\documentclass{aa}  
\usepackage{graphicx}
\begin{document}
\title{Discovery of a planet around the K giant star 4~UMa}


   \author{
	M.P. D\"{o}llinger\inst{1,5}
          \and
	A. P. Hatzes\inst{2}
          \and
	L. Pasquini\inst{1}
          \and
   E. W. Guenther\inst{2}
          \and
	M.~Hartmann\inst{2}
          \and
	L. Girardi\inst{3}
	 \and
	M. Esposito\inst{2,4}
             }

   \offprints{
    Michael D\"{o}llinger \email{mdoellin@eso.org.de}\\$*$~
    Based on observations obtained at the 
    2-m-Alfred Jensch Telescope at the Th\"uringer
    Landessternwarte Tautenburg 
  \\}

     \institute{
European Southern Observatory, Karl-Schwarzschild-Strasse 2, 85748,
Garching bei M\"{u}nchen, Germany
         \and
	Th\"uringer Landessternwarte Tautenburg,
                Sternwarte 5, D-07778 Tautenburg, Germany
		\and 
INAF-Osservatorio Astronomico di Padova, Vicolo dell'Osservatorio 5, I-35122
Padova, Italy
          \and
Dipartimento di Fisica "E.R. Caianiello", Universita di Salerno, via S.
Allende, 84081 Baronissi (Salerno), Italy
          \and
Max-Planck-Institut f\"{u}r Astrophysik, Garching bei M\"{u}nchen, Germany}

   \date{Received; accepted}

\abstract
{
For the past 3 years we have been monitoring a sample of 62 K giant
stars using precise stellar radial velocity measurements with high
accuracy taken at the
Th\"uringer Landessternwarte Tautenburg.}
   {To search for sub-stellar companions to
giant stars and to understand the nature of the diverse
radial velocity  variations exhibited by K giant stars.
   }
   {
We present precise stellar radial velocity measurements
of the K1III giant star 4~UMa (HD 73108).  These were
obtained using the coud{\'e} echelle spectrograph of 2-m Alfred Jensch
Telescope. The wavelength reference for the radial velocity measurements
was provided by an iodine absorption cell.
   }
   {
Our measurements  reveal that the radial velocity  of 4 UMa exhibits
 a periodic variation of 269.3 days with a semiamplitude $K$ $=$
216.8 m\,s$^{-1}$. A Keplerian orbit with an eccentricity, 
$e$ $=$ 0.43 $\pm$ 0.02 is the most reasonable
explanation for the radial velocity variations. The orbit yields a 
mass function, 
$f(m)$ $=$ (2.05 $\pm$ 0.24) $\times$ $10^{- 7}$ $M_{\odot}$.
From our high resolution spectra we calculate a metallicity of $-$0.25 $\pm$ 0.05
and derive a stellar mass of 1.23 $M_{\odot}$ $\pm$ 0.15 for the host star.
}
   {
The K giant star 4 UMa hosts a substellar companion with  minimum mass
$M$~sin~$i$~ = 7.1 $\pm$ 1.6 $M_{Jupiter}$.
}

\titlerunning{Discovery of a planet around the K giant Star 4~UMa}
\maketitle

%
\section{Introduction}

More than 200 extrasolar planets around main sequence stars have been detected
via radial velocity method (RV), however most of these are around
stars with masses of approximately one solar mass. We thus have a very poor
understanding of how the stellar mass influences planet formation, particularly
for more massive stars. Although a few radial velocity searches have been
undertaken for planets around low mass stars (Delfosse et al. 1999;
Endl et al. 2003), there have been fewer attempts
to search for planets around intermediate mass stars. The reason is that
RV searches are ill-suited for early-type, more massive main sequence stars.
These stars are hotter and thus have fewer spectral lines, and these are
broadened significantly by the high rotation rates found among early-type
A-F stars.

In spite of the lower RV precision, there are on-going attempts to search for
planets around intermediate mass A--F stars (Galland et al. 2005a) and
these have produced two substellar companions. A 
9.1 $M_{Jupiter}$ mass companion was found in a 388-d orbit around the
F-type star ($M$ = 1.25 $M_{odot}$) by Galland et al. (2005b). A candidate
brown dwarf (25 $M_{Jupiter}$) in a 28-day orbit was found around the A9V star
HD 180777 (Galland et al. 2006).

An alternative approach to search for planets around more massive stars
is to look at intermediate mass stars that have evolved off the main sequence
and up the giant branch. These stars are cooler and thus have a plethora
of stellar lines. This along with their slower
rotation rates make them amenable to high precision RV measurements. 
Two difficulties are encountered in this
approach. First, unlike for main sequence stars, K giants of widely different
masses can have similar effective temperatures. Second, giant stars show
intrinsic variations due to stellar oscillations.
These short period (2--10 days) variations 
are likely caused by radial and/or nonradial p-mode oscillations 
(e.g. Hatzes \& Cochran 1994).   These oscillations add intrinsic 
RV ``noise'' making the detection of extrasolar planets more difficult.
In spite of these challenges,
searching for planets around giant stars
can give us information about the process of planet formation around
intermediate mass stars.

The first indications of sub-stellar companions around giant stars was found
by Hatzes \& Cochran (1993) who discovered long period RV variations in three
K giant stars. They proposed two viable hypotheses for these variations: 
sub-stellar companions
or rotational modulation. The expected rotational periods of K giants are
several hundreds of days which are comparable to the observed RV periods.
If a large surface inhomogeneity (e.g.  starspot) were on the surface this would
create distortions of the spectral line profiles which would be 
 detected as an RV variation with the rotation period of the star.
For these reasons the nature of the long period RV variations was  not clear.

Subsequent studies have established that giant stars can indeed
host extrasolar planets.
Frink et al. (2002) discovered the first extrasolar planet around the
K giant star HD 137759 ($\iota$ Dra). This was followed
by the discovery of substellar companions to the K2 III
star HD 47536 (Setiawan et al. 2003a) and HD 122430 (Setiawan 2003b).
In the same year Sato et al. (2003) reported a 
 planetary companion around G9 III HD 104985.
Sub-stellar  companions have also been reported for 
HD 11977 (Setiawan et al. 2005) and HD 13189 (Hatzes et al. 2005).
More
recently Hatzes et al. (2006) confirmed  that the initial RV variations
found by Hatzes \& Cochran (1993) in $\beta$ Gem were in fact due to a planetary
companion. This was confirmed by Reffert at al. (2006).

 For most of these discoveries the planet hypothesis for the RV variations
was established due to a lack of variations in other measured quantities 
with the RV period. Spots or stellar pulsations
are expected to also produce variations in either activity indicators (Ca II H\&K), photometry,
and/or spectral line shape variations with the RV period.  These variations
were not found in the giant stars claimed to host extrasolar planets. In the case of $\iota$ Dra
the high eccentricity of the orbit established its Keplerian nature.

\section{Observations and data analysis}

4 UMa (= HD 73108 =  HR 3403 = HIP 42527) belongs to a sample of K giant stars
that we have observed since February 2004
at the Thuringia State Observatory 
(Th\"uringer Landessternwarte Tautenburg or TLS).
Observations were made using the 2m Alfred Jensch Telescope as
part of the Tautenburg Planet Search Program (TOPS).
The TOPS
program uses the coud{\'e} echelle spectrograph  which provides a resolving
power, $R$ = 67000 and  a wavelength range of
4700--7400 Angstr\"{o}ms.
A  total of 46 spectra (nightly averages) were taken of 4~UMa
using the iodine cell.
The total exposure time ranged between 5 and 10~minutes depending on the
weather conditions and this resulted in a typical signal-to-noise ratio 
typically greater 
than 150. The strategy was to make observations for this star 1--3 times
per month, weather permitting. The standard CCD data reductions
(bias-subtraction, flat-fielding and spectral extraction) were performed
using reduction routines of the {\it Image Reduction and Analysis Facility}
(IRAF).

An iodine absorption cell placed
in the optical path provided the wavelength reference for the velocity
measurements.  
The RVs were  calculated by modeling the observed spectra with a high
signal-to-noise ratio template of the star taken without the iodine
cell and a scan of our iodine cell taken
at very high resolution with the Fourier Transform spectrometer (FTS) of the
McMath-Pierce telescope at Kitt Peak.
The relative velocity shift between stellar and iodine absorption
lines as well as the temporal and spatial variations of the instrument profile
were calculated for each observation. 
For more details about the 
spectrograph, RV program, and typical measurement precision
see Hatzes et al. (2005).
For the bright K giant stars in our program we typically achieve
an RV accuracy of about 3--5~m\,s$^{-1}$. A more detailed discussion of the
RV accuracy for the  K giant stars will be presented in a forthcoming paper
on the complete Tautenburg sample.

\section{The properties of the star 4~UMa}
4~UMa has a visual magnitude of $V$ = 5.79 mag and is classified in SIMBAD as 
K1III star. The Hipparcos parallax is 12.92 $\pm$ 0.71 mas and this implies an 
absolute magnitude M$_{\mathrm{V}}$ = 0.146 $\pm$ 0.119 mag. 
The stellar parameters of 4 UMa are summarized in Table 1. These 
were obtained either from the literature or were derived from our analysis of 
the stellar spectra taken without the iodine cell. Our high-quality spectra allowed 
us to determine accurate Fe abundances as well as the effective temperature,
T$_{\mathrm{eff}}$, the surface gravity, $\it{log~g}$,  and the microturbulence 
velocity, $\xi$. The results of this analysis for 4~UMa and the rest
of the Tautenburg sample will be presented in  more
detail in a forthcoming paper.
The metallicity, T$_{\mathrm{eff}}$,
and the absolute V-band magnitude as derived from Hipparcos parallaxes were 
used as input values  
to estimate the mass, age, radius, (B--V)$_{\mathrm{0}}$ and,
in an alternative way, the surface gravity of the program stars 
by comparing these to theoretical isochrones and a 
modified version of 
Jo$\!\!\!/$rgensen \& Lindegren's (2005) method. A detailed description 
about the procedure is given in da Silva et al. (2006).
Previous investigations have also derived stellar parameters
for 4~UMa.  McWilliam (1990)
obtained the following values for the T$_{\mathrm{eff}}$, surface gravity log g
and [Fe/H]: 4370 K, 2.45 and $-$0.26. The corresponding values by Luck (1991)
were 4400 K, 1.61 and $-$0.20, respectively.  Both of these values are in
excellent agreement with our results.
           
\begin{table}[h]
\caption{Stellar parameters of 4 UMa}
\vspace{-0.5cm}
$$
\begin{array}{lll}
\hline
\hline
\mathrm{Spectral\,\,type}	& \mathrm{K1III}      	& \mathrm{HIPPARCOS}\\
m_{V}   			& 5.787\pm0.005   	& \mathrm{mag}	\\
M_{V}     			& 0.146\pm0.119	        & \mathrm{mag}	\\
B-V				& 1.197\pm0.005	        & \mathrm{mag}	\\ 	
\mathrm{Parallax}		& 12.92\pm0.71  	& \mathrm{mas}	\\
\mathrm{Distance}		& 62.39\pm3.43       	& \mathrm{pc}    \\
\mathrm{Mass}^{(a)}		& 1.234\pm0.15	        & \mathrm{M}_{\sun} \\
R_{*}^{(a)} 		        & 18.11\pm1.47	        & \mathrm{R}_{\sun}  \\
\mathrm{Age}^{(a)}	        & 4.604\pm2.0   	& \mathrm{Gyr}  \\
T_{\mathrm{eff}}^{(a)}		& 4415\pm70  		& \mathrm{K}  \\
\mathrm{[Fe/H]}^{(a)}		& -0.25\pm0.04	        & \mathrm{dex}  \\
\log{g}^{(a)}		        & 1.8\pm0.15		& \mathrm{dex}  \\
\mathrm{micro\,turbulence}^{(a)}& 1.2\pm0.8	        & \mathrm{km\,s}^{-1}  	\\
\hline
\hline
\end{array}
$$
{\footnotesize
$^{(a)}$ D\"{o}llinger et al. in preparation}
\end{table}

\begin{figure}[h]
\resizebox{\hsize}{!}{\includegraphics{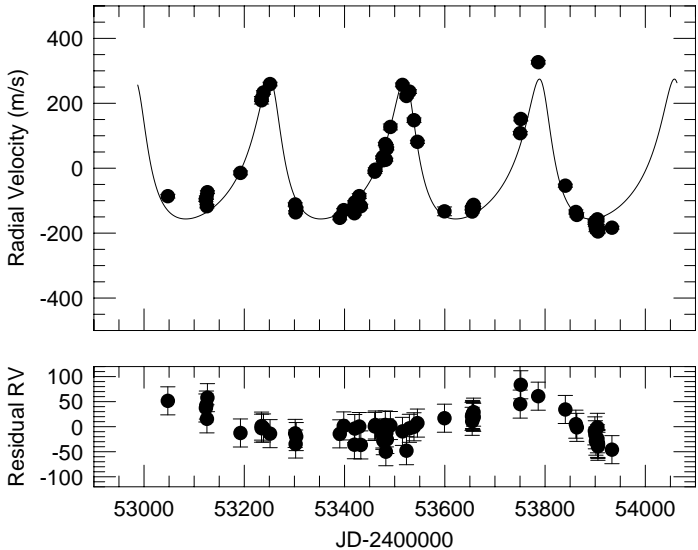}}
\caption{(Top) Radial velocity measurements for 4~UMa. The
solid line is the orbital solution. (Bottom) Residual RV
variations after subtracting the contribution of the planet
orbit.
}
\label{orbit}
\end{figure}

\section{Radial velocity variations and orbital solution}

The time series of our RV measurements for 4 UMa
is shown in Figure 1.
There is an obvious sinusoidal variation in the RV curve with a period
of approximately 270 days. This does not appear to be a pure sine wave and is 
thus the first hint of Keplerian motion.\\

An analysis using a Lomb-Scargle periodogram (Lomb 1976, Scargle 1982)
confirmed the presence of strong power at a frequency
$\nu$ = 0.0038 c\,d$^{-1}$ ($P$ = 271 days). The false alarm probability
(FAP) of this peak using the prescription in Scargle (1982) is estimated to be
FAP $\approx$ 10$^{-8}$.

\begin{figure}[h]
\resizebox{\hsize}{!}{\includegraphics{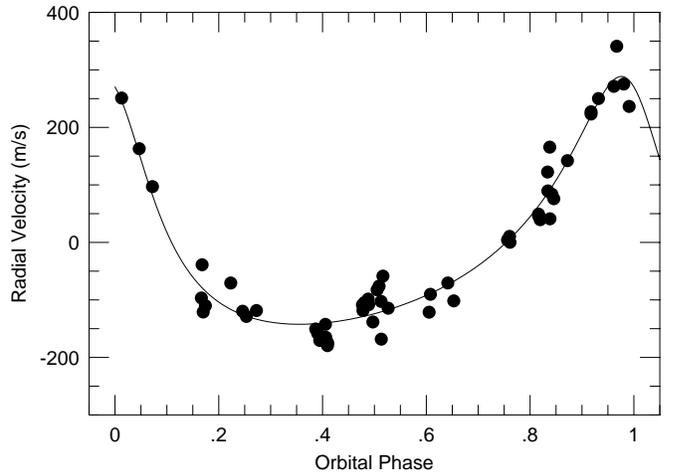}}
\caption{Radial velocity measurements for 4~UMa phased to the
orbital period. The line represents the orbital solution.
}
\label{phase}
\end{figure}

An orbital solution using the RV data resulted in a period, 
P =  269.3 $\pm$ 1.96 days and an eccentricity,  $e$ = 0.43 $\pm$ 0.02.
The corresponding mass function  is
{\it f(m)} = (2.045 $\pm$ 0.247) x~10$^{-7}$~M$_{\odot}$. 
Using our derived stellar mass of $M_{\star}$ = 1.23 $M_{\odot}$ $\pm$ 0.15
we calculated a minimum mass for the companion of
$M$ sin~$i$ = 7.1 $\pm$ 1.6 $M_{Jupiter}$.
All the orbital elements are listed in Table~2. Figure~2
shows the RV variations phase-folded to the orbital period.

\begin{table}
\caption{Orbital parameters for the companion to 4~UMa.}
\begin{tabular}{ll}
\hline
\hline
Period [days]                   & 269.3 $\pm$ 1.96  \\
T$_{periastron}$[JD]            & 52987.3936 $\pm$ 4.31 \\
K[ms$^{-1}$]                    & 215.55 $\pm$ 7.10  \\
$\sigma$(O--C)[ms$^{-1}$]     & 28.8   \\ 
e                               & 0.432 $\pm$ 0.024  \\
$\omega$[deg]                   & 23.81 $\pm$ 4.42 \\
f(m)[solar masses]              & (2.045 $\pm$ 0.247) x 10$^{-7}$ \\
a[AU]                           & 0.87 $\pm$ 0.04 \\ 
\hline
\hline
\end{tabular}
\end{table}

Although the orbital fit to the data is good, there are points that 
significantly from the solution. In fact, the rms scatter,
$\sigma$, about the orbital solution is rather large,
$\sigma$ $\approx$ 30 m\,s$^{-1}$, or about a factor of ten  larger
than our expected errors. We suspect that this results from another 
periodic signal in the data. The lower panel of Figure 1 shows
the residual RV variations after removal of the orbital contribution due 
to the planetary companion.  There are clear variations with a period
much larger than the orbital period. This is confirmed by the
Lomb-Scargle periodogram of the radial velocity residuals  shown in 
Figure 3.  This shows a strong peak at a frequency of $\nu$ = 
0.00156 c\,d$^{-1}$ ($P$ = 641 days). The statistical significance
of this signal was estimated using a `bootstrap randomization' technique.
The measured RV values were randomly shuffled
keeping the observed times fixed and a periodogram for the shuffled data
computed. After 2$\times$ 10$^{5}$ shuffles there was no instance where the
random fake data periodogram data had higher power than the data periodogram in the
frequency range. This indicates that the FAP $<$ 2 $\times$ 10$^{-6}$. This
signal is thus statistically significant.

We also note that K giants show RV variations on time scales of several days and amplitudes of 
up to 100 m\,s$^{-1}$.  This may also contribute to the 
scatter of the RV points about the orbital solution.

\begin{figure}[h]
\resizebox{\hsize}{!}{\includegraphics{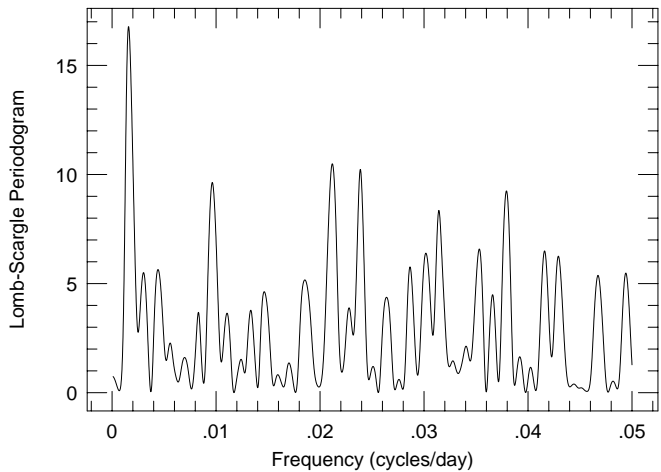}}
\caption{Lomb-Scargle periodogram of the RV residuals of 4 UMa. The strong
peak corresponds to a period of 641 days.}
\label{ftres}
\end{figure}

\section{Hipparcos photometry}

We also analyzed the Hipparcos photometry of  4 UMa in order to see if 
variations with the 270 day period are present. 
Hipparcos obtained 27 observations (`daily" averages') over a 3 year
period.
Figure 4 shows the Lomb-Scargle
periodogram of the Hipparcos photometry after averaging the multiple
measurements taken within a few hours of each other. There was on obvious
outlier that had a value more than 10$\sigma$ from the mean
that was eliminated from the data prior to calculating the periodogram.
Although there is a weak peak near the orbital frequency this is not 
significant and there are several peaks with more power. The highest peak
is at a frequency of 0.0252 c\,d$^{-1}$ ($P$ = 39.698 days). Figure 5 shows
the photometry phased to this period. The false alarm probability 
was estimated using the bootstrap randomization technique.
After 100,000
such shuffles 19\% of the random data periodograms showed power greater
than the real periodogram. This peak is most likely not significant.
It it were real it be could be due to stellar oscillations. 
For example, Hatzes \& Cochran (1998) detected a similar
period ($\approx$ 50 days) in the spectral line bisector variations in Aldebaran
which they attributed to oscillations.
 
\begin{figure}[h]
\resizebox{\hsize}{!}{\includegraphics{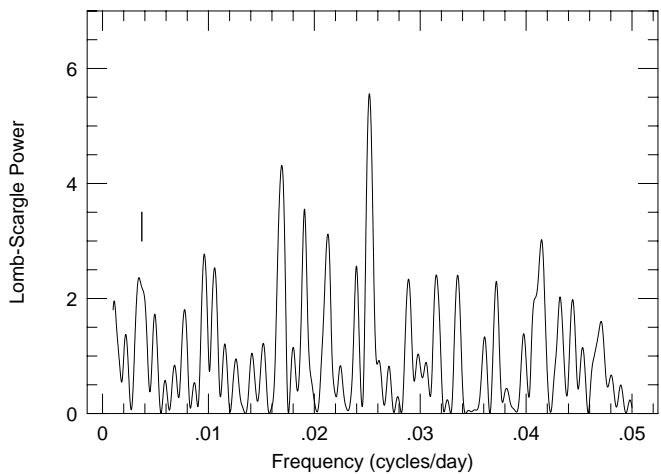}}
\caption{Lomb-Scargle periodogram of the Hipparcos photometry for
 4~UMa. The horizontal line marks the orbital frequency.}
\label{phot}
\end{figure}

\section{Discussion}

Our radial velocity measurements indicate that the
giant star 4 UMa hosts a giant extrasolar planet.
We are confident that these RV variations
are due to a sub-stellar companion and not rotational modulation
for three reasons. First, the orbit is
highly eccentric and such a saw-toothed RV pattern is difficult
to reproduce with rotational modulation from spots or possibly even
from stellar oscillations. Indeed, the high eccentricity of the companion
to $\iota$ Dra was the convincing argument that this was a true companion.
Second, the Hipparcos photometry does not show any significant variations
at the orbital period. Finally, our periodogram analysis shows a statistically
significant period at 641 days in the RV residuals for 4 UMa. We have
detected two periods of several hundred days; both cannot be due to rotation.
Because the RV variations of the dominant 269-d RV period is well-fit by an 
eccentric Keplerian orbit, we believe that this signal is due to a planetary companion.
The 641-day period we have detected may indeed be due to rotational modulation,
although a similar signal does not appear in the Hipparcos photometry.
We caution the reader, however, that the Hipparcos photometry was not
contemporaneous with our RV measurements.

\begin{figure}[h]
\resizebox{\hsize}{!}{\includegraphics{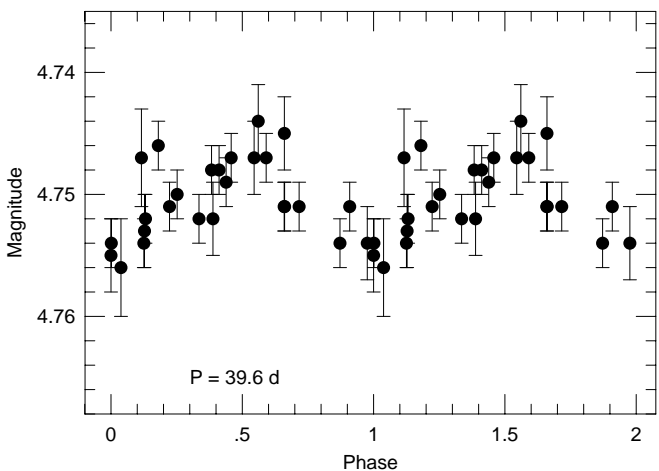}}
\caption{Hipparcos photometry for 4~UMa phased to the
39.6 day photometric period. }
\label{orbit}
\end{figure}

At this point we do not wish to speculate as to the nature of the 641-day
period that we have detected in the RV residuals. It could well be due
to rotational modulation, or possibly due to a second companion. We are
continuing to monitor this star and are currently making a detailed analysis
of other observed quantities (H$\alpha$ for example) to see if this in fact 
the rotation period of the star.

The high eccentricity of the orbit with a value of $e$ = 0.43 $\pm$ 0.02 is 
not unusual for planet hosting stars, including giant stars.
The companion to $\iota$ Dra has the highest eccentricity ($e$ = 0.7)
for a planet around a giant star. 
Relatively high eccentricities have also been found
for the companions to HD 11977 ($e$ = 0.4, Setiawan et al. 2005) and HD 13189 
($e$ = 0.27, Hatzes et al. 2006).
It seems that the planetary companions to giant stars can have the wide
range of orbital eccentricities that are shown by planets orbiting main
sequence stars.

The abundance analysis for 4~UMa shows that this star is slightly metal poor
([Fe/H] = $-$0.25 $\pm$ 0.05).
This  interesting because this the main sequence host stars of 
exoplanets tend to to be metal rich compared to stars that do not 
posses exoplanets (Santos et al. 2004). 
Other authors (Schuler et al. 2006; da Silva et al. 2006) have also
found evidence that  planet hosting giant stars are metal poor.
However, given the small number of planets around giant stars it is premature
to conclude that these contradict the planet-metallicity trend for
main sequence stars. An analysis of a larger sample of giant stars may
show that those have higher metallicity indeed have a higher frequency
of sub-stellar companions. 

The number of exoplanets around evolved stars is so far very limited.
The discovery of the companion to 4 UMa adds to the growing list of planets
evolved stars hosting sub-stellar companions
(see Hatzes et al. 2006 and references therein). The estimated  masses for
the giant stars hosting planets range from 1 to 3 solar masses. In terms
of the properties of planets around giant stars they have similar 
characteristics, masses in the range of  of 2--14 $M_{Jupiter}$ and orbital
periods of several hundred days (see discussion by Hatzes et al. 2006). In 
other words, similar to the properties of the planet around 4~UMa.
These discoveries are important because they probe a different
stellar mass  regime different from the main sequence objects that are the
targets of most planet searches.

\begin{acknowledgements}
We are grateful to the user support group of the Alfred-Jensch telescope:
B. Fuhrmann, J. Haupt, Chr. H\"{o}gner, U. Laux, M. Pluto, J. Schiller,
and J. Winkler. We thank Luciano Fraga for taking some of the observations.
This research has made use of the SIMBAD data base operated
at CDS, Strasbourg, France.
\end{acknowledgements}


\begin{thebibliography}{}
\bibitem[da Silva et al. 2006]{da} da Silva, L., Girardi, L., Pasquini, L.,
 Setiawan, J., von der L\"uhe, O., de Medeiros, J.R.,
Hatzes, A.P., D\"ollinger, M.P., Weiss, A. 2006, A\&A in press
\bibitem[Delfosse et al. 1997]{del97} Delofosse, X., Forveille, T., 
Beuzit, J.-L., Udry, S., Mayor, M., \& Perrier, C. 1999, A\&A, 344, 897.
\bibitem[Endl et al. 2003]{endl03} Endl, M., Cochran, W.D., Tull, R.G., \&
MacQueen, P.J. 2003, AJ, 3099.
\bibitem[Frink et al. 2002]{frink02} Frink, S., Mitchell, D.S., Quirrenbach, A., et al. 2002, ApJ, 576, 478
\bibitem[Galland et al. 2005a]{gall05a} Galland, F., Lagrange, A.-M., Udry, S., 
Chelli, A., Pepe, F., Queloz, D., 
Beuzit, J.-L., Mayor, M. 2005a, A\&A, 443, 337.
\bibitem[Galland et al. 2005b]{gall05b} Galland, F., Lagrange, A.-M., Udry, S., 
Chelli, A., Pepe, F., Beuzit, J.-L., Mayor, M. 2005b, A\&A, 444, 21.
\bibitem[Galland et al. 2006]{gall06} Galland, F., Lagrange, A.-M., Udry, S., 
Beuzit, J.-L., Pepe, F., Mayor, M. 2006, A\&A, 452, 709.
\bibitem[Hatzes \&\ Cochran1993]{hatzes93} Hatzes, A.P., Cochran, W.D.
 1993, ApJ, 413, 339
\bibitem[Hatzes \&\ Cochran1994]{hatzes94} Hatzes, A.P., Cochran, W.D.
 1994, ApJ, 422, 366
\bibitem[Hatzes \&\ Cochran1998]{hatzes98} Hatzes, A.P., Cochran, W.D.
1998, MNRAS, 293, 469
\bibitem[Hatzes et al. 2005]{hatzes05} Hatzes, A.P., Guenther, E.W., Endl, M.,
 et al. 2005, A\&A, 437, 743
\bibitem[Hatzes et al. 2006]{hatzes06} Hatzes, A.P., Cochran, W.D.,
 Endl, M. et al.  2006, A\&A, 457, 335
\bibitem[]{jor05}  Jo{$\!\!\!/$}rgensen, B.R., Lindegren, L. 2005, A\&A, 436, 127
\bibitem[Lomb1976]{lomb76} Lomb, N.R. 1976, Ap\&SS, 39, 477
\bibitem[Luck1991]{luck91} Luck, R.E. 1991, ApJS, 75, 579--610
\bibitem[McWilliam1990]{mcwilliam90} McWilliam, A. 1990, ApJS, 74, 1075M
\bibitem[Reffert et al. 2006]{reffert06} Reffert, S., Quirrenbach, A.,
 Mitchell, D.S., Albrecht, S., Hekker, S.,
Fischer, D., Marcy, G.W., Butler, R.P. 2006, ApJ, 652, 661.
\bibitem[Santos et al. 2004]{santos04} Santos, N.C., Israelian, G., Mayor, M.
 2004, A\&A, 415, 1153
\bibitem[Sato2003]{sato03} Sato, B., Ando, H., Kambe, E. 2003, ApJ, 597L,
 157
\bibitem[Scargle1982]{scargle92} Scargle, J.D. 1982, ApJ, 263, 835
\bibitem[Schuler2006]{schuler06} Schuler, S., Kim, J.H., Tinker, M.C., Jr., 
King, J.R., Hatzes, A.P., Guenther, E.W. 2005, ApJ, 632, L131.
et al. 2006, 
\bibitem[Setiawan et al. 2003a]{setiawan03a} Setiawan, J., Hatzes, A.P., von 
der L\"{u}he, O., et al. 2003a, A\&A, 398, L19
\bibitem[Setiawan 2003b]{setiawan03b} Setiawan, J., Pasquini, L., da Silva, L.,
et al. 2003b, A\&A, 397, 1151S
\bibitem[Setiawan et al. 2005]{setiawan05} Setiawan, J., Rodman, J., da
 Silva, L., et al. 2005, A\&A, 437L, 31     
\end{thebibliography}
\end{document}